\documentclass[review, 12pt, numbers, sort&compress]{elsarticle}

\usepackage{changes}
\usepackage{amsmath}
\usepackage{wasysym}
\usepackage{amssymb}  %数学符号
\usepackage{amsfonts}  %数学公式、矩阵、积分求和等
\usepackage{multirow}
\usepackage{txfonts}  %默认字体 新罗马

\usepackage{textcomp}
\usepackage{stfloats}
\usepackage{url}
\usepackage{verbatim}
\usepackage{graphicx}
\usepackage{dsfont}
\usepackage{multirow}
\usepackage{color}
\usepackage{caption}  % 改变图标标题

\usepackage{geometry}
\usepackage[font=small,labelfont=bf,labelsep=none]{caption}  %加粗
\usepackage{balance}

\usepackage{setspace}
\usepackage[T1]{fontenc} % 保证英文字体加粗有效
\usepackage{booktabs}    %调整表格线与上下内容的间隔
\usepackage{algorithm}  
\usepackage{algorithmic}   
\usepackage{float}
\usepackage{epstopdf}
\usepackage{subfigure}
\RequirePackage{caption}

\newtheorem{lemma}{Lemma}

\newtheorem{proposition}{Proposition}
\newtheorem{definition}{Definition}
\newtheorem{proof}{Proof}



\usepackage{siunitx}
\usepackage{bm}
\usepackage{array}  
\makeatletter
\newcommand\multiline[1]{\parbox[t]{\dimexpr\linewidth-\ALG@thistlm}{#1}}
\makeatother
\usepackage{calligra}

\PassOptionsToPackage{noend}{algpseudocode}% comment out if want end's to show
\def\dif{\mathop{}\hphantom{\mskip-\thinmuskip}\mathrm{d}}%
\let\daccent\d
\let\d\relax
\newcommand\d{\ifmmode\dif\else\expandafter\daccent\fi}

\usepackage[english]{babel}   %按音节换行
\allowdisplaybreaks[4]   %允许公式跨页
\usepackage{etoolbox}
\makeatletter
\patchcmd{\@makecaption}
{\scshape}
{}
{}
{}
\makeatletter
\patchcmd{\@makecaption}
{\\}
{.\ }
{}
{}
\makeatother

\errorcontextlines\maxdimen

\journal{Signal Processing}

\begin{document}
	
\begin{frontmatter}  %不计页数的部分

\title{\textcolor{blue}{Augmented LRFS-based Filter: Holistic Tracking of  Group  Objects}}

\author[SEU]{Chaoqun Yang\corref{cor1}}
\ead{ycq@seu.edu.cn}
\cortext[cor1]{Corresponding Author}

\author[SEU]{Xiaowei Liang}
\ead{ lxwysa@126.com}

\author[ZJU]{Zhiguo Shi}
\ead{shizg@zju.edu.cn}

\author[JOU]{Heng Zhang}
\ead{zhangheng@jou.edu.cn}

\author[SEU]{Xianghui Cao}
\ead{xhcao@seu.edu.cn}

\address[SEU]{School of Automation, Southeast University, Nanjing, 210096, China}
\address[ZJU]{College of Information Science and Electronic Engineering, Zhejiang University, Hangzhou, 310027, China}
\address[JOU]{School of Computer Engineering, Jiangsu Ocean University, Lianyungang, 222000, China}

\begin{abstract}
\textcolor{blue}{Aiming at the problem of accurate tracking of group objects,
where} multiple closely spaced objects within a group pose a coordinated motion,
\textcolor{blue}{this paper develops a new type  of  labeled random finite set (LRFS), i.e., augmented LRFS,
which inherently integrates  group information such as the group \textcolor{blue}{geometry} center and the group index
into the definition of LRFS.}
Specifically, for each element  in an augmented LRFS, the kinetic states,  the track labels, and the corresponding group information of   its represented object are incorporated.
Then, \textcolor{blue}{by means of the  proposed augmented LRFS-based filter, i.e., }
the labeled multi-Bernoulli   filter with the proposed augmented LRFS, 
the group structure is iteratively propagated and updated during the tracking process,
which achieves the \textcolor{blue}{holistic estimation} of the kinetic states, track labels, and the corresponding group information of multiple group objects, and further improves the  tracking performance.
\textcolor{blue}{Finally, simulation experiments are conducted to verify 
the  effectiveness of the  proposed augmented LRFS-based filter.  }
\end{abstract}

\end{frontmatter}

\section{Introduction}
{N}{owadays},
the issue of group object tracking (GOT) has attracted increasing interest due to its   wide  applications over civil and military fields,
such as \textcolor{blue}{drone swarms~\cite{Guerlin20swarm,Longliu}, } vehicle formations~\citep{Riuiping,Wang2018probabi,8805104}, and group of robots~\citep{Jointt2021,Senanayake2016sear,BoxParticle}, etc.
In these applications, multiple objects such as drones or vehicles,
are usually closely spaced and evolve in a coordinated manner with the same or similar dynamics models, forming one or many groups.
Moreover,  \textcolor{blue}{as shown in Fig.~1,} these groups can split, merge, die and rebuild~\citep{zhang2023groupi},
performing complex dynamics of group structure and frequency interaction with the group objects within or without groups.
Thus, the GOT issue not only suffers from the difficulties encountered by the issue of multi-object tracking (MOT) including
unknown number of objects,  missed detections, clutters, and measurement origin uncertainty,
but also  encounters the indeterminacy of group structure caused by group formation, death, merging or splitting~\citep{zhang2022Seamless,7012075,Xianpengq1}.

\begin{figure}
	\centering
	\includegraphics[width=1\textwidth]{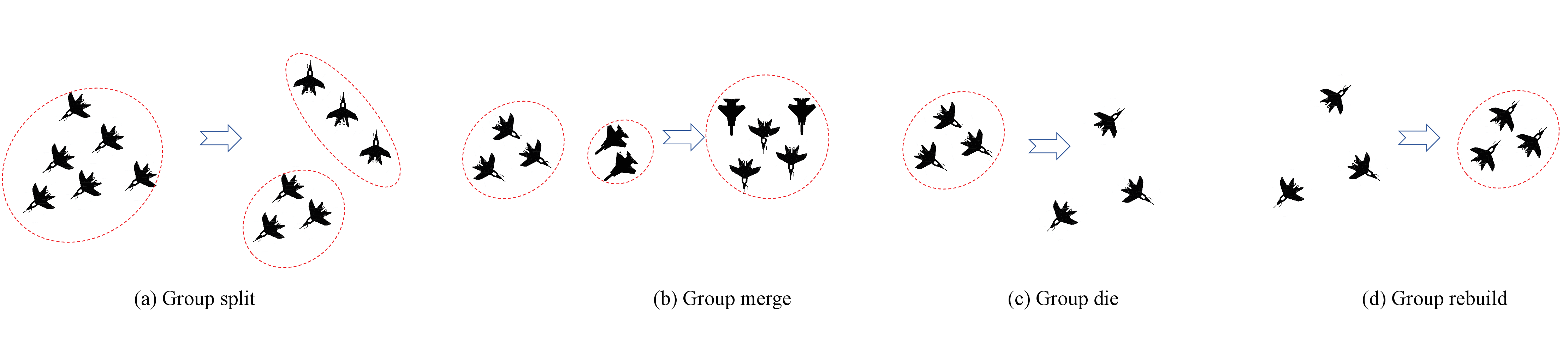}\\
	\caption{{\textcolor{blue}{The diagram of the dynamics  of  group structure, including group split, merge, die and rebuild.}}}
\end{figure}

Generally speaking,  most of the current GOT methods can be \textcolor{blue}{divided} into three categories: 
\textcolor{blue}{unresolvable GOT  based methods},  resolvable GOT (RGOT) based methods, and partially resolvable GOT based methods.
The first takes each group as an extended object,
and pays more attention to the estimation of each group's   center and shape,
while the second focuses on the estimation of both the states of multiple objects and group structures~\citep{Mihaylova2014Monte,zhang2023groupi}.
\textcolor{blue}{In the scenario targeted by the third category, resolvable and unresolvable group objects coexist~\cite{Longliu}. This scenario aligns better with real-world applications.
However, it is also more difficult and challenging to deal with this scenario because it usually requires simultaneous tracking of both resolvable and unresolvable  group objects. Unfortunately, there is relatively little research on the third category.}

\textcolor{blue}{Aiming} at the estimation of the time-varying number of objects and groups, multiple objects,
as well as  group structure,
in this paper, the RGOT methods are considered.
Depending on whether random finite set (RFS) theory is used  or not,
the main RGOT based methods can be \textcolor{blue}{divided} into two kinds,  data association (DP) based methods and RFS-based methods.

For the  DP-based RGOT methods,
Zhang et al.  considered the GOT problem within the multiple hypothesis tracking (MHT) framework,
and proposed a  MHT-based method to efficiently deal with the  problems of group association, group partition and  the state estimation of the individual objects within groups~\citep{zhang2023groupi}.
Gning et al. combined evolving graph network,  sequential Monte Carlo method, and joint probabilistic data association methods  to cope with the GOT issue~\citep{Gning2022group}.
Zhang et al. utilized the belief propagation method, also  known as the message passing method, and addressed  a belief propagation method to  jointly
estimate  both object states and group structure~\citep{zhang2022Seamless}.

Due to the advantages of coping with the uncertainty of both the number and states of multiple objects,
RFS-based  RGOT methods, such as   labeled multi-Bernoulli  (LMB) filter, multi-Bernoulli filter, Poisson multi-Bernoulli mixture (PMBM) filter,
have received much more attention than the DP-based \textcolor{blue}{methods~\citep{liu2018structure,Li2022resolve}.}
For example, Li et al. proposed the leader follower labeled multi-Bernoulli  filter which combined the  LMB filter with leader follower (LF) model~\citep{Li2023labeld}.
Thereinto, the LF model was used to formulate the dynamics of group structure, and the LMB filter was regarded as the filtering tool.
Liu et al.  classified the GOT issue into two steps: object state estimation and group state including group structure, group size, etc, estimation,
of which the former utilized the multi-Bernoulli filter, and the latter was based on graph theory~\citep{liu2018structure}.
Similarly, Zhou et al.  also adopted graph theory to describe group structure,
and used virtual leader follower model to predict the evolution of groups,
then, the PMBM filter was used to track multiple resolvable group   objects~\citep{zhou2024poisson}.
Li et al. proposed the single object state transtition function based multi-Bernoulli filter and studied its application to the sensor control problem in the RGOT scenarios~\citep{Li2022resolve}.

For the RFS-based methods, although the above  existing  remarkable works have presented diverse methods on how to integrate the group structure into the RFS-based filters
to furthe improve the tracking performance,
it is worth noting that in the processing of these methods, the multi-object states and group structures are still treated independently. They are processed separately rather than considered as a unified whole.
Specifically, these methods formulate the multi-object state as an RFS without incorporating group structure information into the RFS. Consequently, the estimation of multi-object states and the estimation of group structures are two independent processes. In fact, the estimation of group structures often serves as a post-processing step after the estimation of multi-object state.
Therefore, a natural question arises: Can the RFS theory naturally integrate group structure information and achieve a holistic estimation of both object states and group structures? To the best of our knowledge, there is currently no work in this direction.

Inspired by the advance of labeled RFS (LRFS),
which is a special kind of RFS that jointly integrates object states  and object labels,
in this paper, we propose a new kind of LRFS, named augmented  LRFS,
that simultaneously integrates the information of object states,  labels, and group structure.
Subsequently, the statistical characteristic of the proposed augmented LRFS is analyzed in details.
Furthermore, based on the  proposed augmented LRFS, 
we formulate the GOT problem within the framework of the augmented LRFS based multi-object Bayes filter,
and propose an augmented LRFS-based  filter named   the LMB filter with the augmented LRFS to achieve the accurate tracking of group objects.
By means of the proposed filter, both the states and labels of multiple objects, along with group structures are simultaneously
estimated.
In summary, the main contributions of this paper can be listed as follows.
\begin{itemize}
	\item We propose a new kind of RFSs, named augmented LRFS,
	which integrates group information into the existing LRFS, and we further analyze  the statistical characteristic of the proposed augmented LRFS  in details.
	\item We propose a new formulation of the GOT issue based on the proposed augmented LRFS,
	in which  the dynamics of both multiple objects and groups are integrated into the augmented LRFS-based multi-object Bayes filter.
	\item We propose a new method to cope with the GOT issue, i.e.,  
	an augmented LRFS-based  filter named  LMB filter with the augmented LRFSs,
	and prove its feasibility to solve the GOT issue.
	As far as we know, it is the first RFS-based method that naturally integrates group structure information and achieves a holistic estimation of both object states and group structures.
	
\end{itemize}

The rest of this paper is organized as follows: Section 2
presents the proposed augmented LRFS theory. 
Section 3 states the considered GOT problem.
Section 4 describes augmented LRFS-based  filter that  copes with the GOT problem,
Numerical experiments are provided in section 5.
Finally, section 6 concludes this paper.

\section{Proposed  Augmented LRFS}
This section will present the definition and statistical properties of the  proposed augmented  labeled LRFS.
%after a brief review on the concept of both random finite sets (RFSs) and labeled random finite sets (LRFSs).
\textcolor{blue}{Before that, the notations and operators used in the rest of this paper are described in Table 1.}
\begin{table}[h!]
	\begin{center}
		\caption{ \textcolor{blue}{The description of notations and operators.} }
		\label{iou1}
		%		\begin{tabular}{  p{150pt}   c  c c  }%{\linewidth}
			\begin{tabular*}{\textwidth}{@{\extracolsep{\fill}} | l| l |} 
				\hline
				\textbf{Notations/Operators}  &\textbf{Description/Definition}\\% & {Medium level}   \\ 
				\hline
				Small English letters (e.g., $x$)    & Random variables  \\ %&\SI{5}{m}   \SI{8}{m}  \SI{13}{m} \\ 
				\hline
				Capital English letters (e.g., $X$)  & RFSs  \\%&3(d)  3(e) 3(f) % \\ 
				\hline
				Bold face English letters  (e.g., $\mathbf{X}$)   & LRFSs   \\
				  \hline
				Capital English letters written with scans serif font (e.g., $\mathsf{X}$)     & Augmented LRFSs     \\ %  &0.55  &   0.74 &     0.81   \\ 
				  \hline
			 Blackboard bold English letter   (e.g., $\mathbb{X}$, $\mathbb{L}$)   & Spaces  (e.g.,state space, label space)    \\% &0.34  &   0.33  &   0.29   \\ 
				\hline
				$||\cdot||_2$  &2-norm  \\
					\hline
					$<f,g>=\int f(x)g(x)dx$  &Inner product\\
						\hline
				$h^X=\prod_{x\in X}h(x)$ 	&RFS exponential operator \\
					\hline
				$\delta_Y(X)$	&Kronecker delta function~\citep{Vo2014labeled}. \\
					\hline
				$1_Y(X)$	&Inclusion fuction~\citep{Vo2014labeled}.\\
					\hline
			\end{tabular*}
		\end{center}
	\end{table}

%\subsection{Notation}
%In the rest of this paper, small English letters (e.g., $x$) are used to denote random variables.
%Capital English letters (e.g., $X$) are used to denote RFSs  (see section 2).
%Bold face English letters  (e.g., $\mathbf{X}$) are used to represent LRFSs (see section 2.3).
%Capital English letters written with scans serif font (e.g., $\mathsf{X}$) are used to represent augmented LRFSs (see section 2.4).
%Additionally, spaces are denoted by blackboard bold English letter, for example, 
%$\mathbb{X}$ denotes the state space,
%$\mathbb{L}$ denotes the label space,
%$\mathbb{N}$ denotes the integer  space,
%$\mathbb{Z}$ denotes the natural number space.
%
%The following operators are defined. $||\cdot||_2$ denotes the 2-norm, 
%$<f,g>=\int f(x)g(x)dx$ denotes the inner product.
%For an RFS $X$, we define the following exponential operator,
%\begin{equation}
%	h^X=\prod_{x\in X}h(x).
%\end{equation}
%In addition, similiar to~\citep{Vo2014labeled}, we also use
%\begin{eqnarray}\label{nonn}
%	\delta_Y(X)= \left\{\begin{array}{ll}
%		1, & \mbox{if $X=Y$}, \\
%		0, & \mbox{otherwise},
%	\end{array} \right.
%\end{eqnarray}
%and 
%\begin{eqnarray}\label{nonn11}
%	1_Y(X)= \left\{\begin{array}{ll}
%		1, & \mbox{if $X\subseteq Y$}, \\
%		0, & \mbox{otherwise},
%	\end{array} \right.
%\end{eqnarray}
%to denote the Kronecker delta function and the inclusion fuction, respectively.

\subsection{ RFSs $\&$ LRFSs}
\begin{definition}
	An RFS like $X=\{x_1,x_2,\cdots, x_n\}$, is essentially a finite-set-value random variable,
	in which both the number of elements and all of the  elements are random~\citep{Yang2023distri,Wen1}.
%	\begin{itemize}
%		\item  Each element $x_i\in X$ is random and unordered;
%		\item  The cardinality (i.e., the number of elements) of $X$, $|X|=n$ is random.
%	\end{itemize}
\end{definition}

\textcolor{blue}{According to Finite Set Statistics (FISST)~\citep{Mahler2007s5a5,Jun1}, the statistical characteristics of  an RFS $X$ can be fully captured by its corresponding FISST probability density function (PDF) $\pi(X)$. 
There are many types of RFSs, such as Possion RFS, Bernoulli  RFS and Multi-Bernoulli  RFS, etc,
their definition can be found in Ref.~\citep{Vo2013labeled,Mahler2007s5a5,Reuter2014label,IMMLMB,Jun2}.}

%Specially, we introduce the following two types of RFSs that will be used in the subsequent sections.
%\subsubsection{Possion RFS}
%A Possion RFS is an RFS  $X$ whose PDF follows~\citep{Vo2013labeled}
%\begin{equation}
%	\pi(X)=e^{-<v,1>}v^X,
%\end{equation}
%where $v$ is a given function (also called intensity function).
%
%\subsubsection{Bernoulli  RFS}
%A Bernoulli RFS is an RFS  $X$ whose PDF follows~\citep{Vo2013labeled} 
%\begin{eqnarray}
%	\pi(X)= \left\{\begin{array}{ll}
%		1-r, & \mbox{$X=\emptyset$}, \\
%		r\cdot p(x), & \mbox{$X=\{x\}$},
%	\end{array} \right.
%\end{eqnarray}
%where  $p$ and $r$  are a PDF and the  existence probability of $X$, respectively.
%\subsubsection{Multi-Bernoulli  RFS}
%By taking the union of a certain number of mutually independent Bernoulli RFSs, 
%i.e., $X=\cup_{i=1}^M{X^{i}}$, we further construct a multi-Bernoulli  RFS, whose
%PDF follows \citep{Reuter2014label,IMMLMB}
%\begin{small}
%	\begin{equation}
%		\pi(\{x_1,\cdots,x_n\})=\prod_{j=1}^M(1-r^{(j)})\sum_{i\leq i_1\neq \cdots \neq i_n \leq M}\prod_{j=1}^n 
%		\frac{r^{(i_j)}p^{(i_j)}(x_j)}{1-r^{(i_j)}}.
%	\end{equation}
%\end{small}
%
%For simplicity, we usually use $\pi=\{(r^{(i)},p^{(i)})\}_{i=1}^M$ to abbreviate the above PDF. 

%\subsection{ LRFSs}
\textcolor{blue}{By augmenting the  state $x$ with a distinct label $l\in \mathbb{L}$, we have $\mathbf{x}=(x,l)$.
In the context of object tracking, specifically, $l$ can be represented as $l=(k,i)$, where $k$ is 
the object's birth time, and $i$ is a unique identifier used to distinguish objects borning  at the same time. In the following context, unless otherwise stated, $l=(k,i)$ holds true.
Further,  we can construct a new RFS $\mathbf{X}=\{\mathbf{x}_1,\mathbf{x}_2,\cdots, \mathbf{x}_n\}$.
If the labels of  all of the elements in $\mathbf{X}$ are distinct, we call the RFS $\mathbf{X}$ as an LRFS.
Then we obtain the following definition of LRFSs.
\begin{definition}
	An LRFS with state space  $\mathbb{X}$ and  label space  $ \mathbb{L}$, like  $\mathbf{X}=\{\mathbf{x}_1,\mathbf{x}_2,\cdots, \mathbf{x}_n\}$,  is an  RFS on   $\mathbb{X}\times \mathbb{L}$  such that each realization has distinct labels~\citep{Vo2014labeled}. 
\end{definition}}

\textcolor{blue}{There are also some types of LRFS, such as labeled multi-Bernoulli (LMB) RFS
and generalized LMB  (GLMB) RFS.
For instance, the LMB RFS has the following compact PDF form \citep{Reuter2014label,Shi1,Chen2}}

\textcolor{blue}{\begin{equation}\label{lmbeqa}
	\pi(\mathbf{X})=\bigtriangleup(\mathbf{X})\omega(\mathcal{L}(\mathbf{X}))p^{\mathbf{X}}
\end{equation}
where 
\begin{eqnarray}
	\omega(L) &=& \prod_{j\in \mathbb{L}}(1-r^{(j)})\prod_{l\in L}\frac{1_{\mathbb{L}}(l)r^{(l)}}{1-r^{(l)}},\\
	p(x,l)&=&p^{(l)}(x),
\end{eqnarray}
 $\mathcal{L}(\cdot)$ denotes the projection from $\mathbb{X}\times \mathbb{L}$ to $ \mathbb{L}$, 
$p^{l}(x)$ and $r^{l}(x)$  are the PDF and the  existence probability of $x$ with label $l$, respectively,
and $
	\bigtriangleup({\cdot})=\delta_{|\cdot|}(\mathcal{L}(|\cdot|))$
is a distinct label indicator to ensure distinct labels. 
For simplicity, we usually use the parameter set $\pi=\{(r^{(l)},p^{(l)})\}_{l \in \mathbb{L}}$ to abbreviate the above PDF. }

%\subsubsection{Generalized Labeled Multi-Bernoulli  RFS}
%A GLMB RFS  is  an LRFS with the following PDF~\citep{{Vo2013labeled}} 
%\begin{equation}
%	\pi(\mathbf{X})=\bigtriangleup(\mathbf{X})\sum_{c\in \mathbb{C}}{\omega^{(c)}(\mathcal{L}(\mathbf{X}))[p^{(c)}]^{\mathbf{X}}}
%\end{equation}
%where   
%$\mathbb{C}$ is a discrete index set, and 
%\begin{eqnarray}
%	\sum_{L \subseteq \mathbb{L}}\sum_{c \subseteq \mathbb{C}}	{\omega(L)}&=&1,\\
%	\int p^{(c)}(x,l)dx&=&1.
%\end{eqnarray}

\subsection{Augmented LRFS}
To address the GOT issue, in this paper, 
we propose the concept of  augmented LRFS, which can be taken as the extension of the concept of LRFS.

Firstly,  for the label of each object,  defined in an  LRFS, $l=(k,i)$, 
we augment it by integrating the group information of the group that the object belongs to.
Specifically, an augmented label is defined as 
\begin{equation}\label{alrfsxx}
	\rho=(l,g,c)=(k,i,g,c)
\end{equation}
where $l=(k,i)$,
and 
\begin{itemize}
	\item $k$ still denotes the birth time of this object.
	\item $i\in \mathbb{N} $  still denotes the distinct indice to distinguish the objects  born at the same time.
	\item $g\in \mathbb{Z} $ denotes the unique index of  the group that the object belongs to.  Specially, $g=0$ denotes the object does not belong to any group.
	\item ${c}\in\mathbb{X} $ denotes the \textcolor{blue}{geometry} center  of the group $g$.
\end{itemize}

Secondly, for $\mathsf{x}=(x,\rho)$, let $\mathbb{L}^1, \mathbb{L}^2,\mathbb{L}^3,\mathbb{L}^4$ denote the spaces to which $k,i,g,{c}$ belong, respectively.
Let $\mathbb{L}^{m}\times \mathbb{L}^{n}$ denotes the space to which the $m$-th element and  $n$-th element jointly belong.
For example, $\mathbb{L}^{3}\times \mathbb{L}^{4}$ denotes the space to which $g$ and  $c$ jointly belong.
Let $\mathbb{L}=\mathbb{L}^1 \times  \mathbb{L}^2 \times \mathbb{L}^3 \times \mathbb{L}^4$,
then, we have $l\in \mathbb{L}^{1}\times \mathbb{L}^{2}$, $\rho\in \mathbb{L}$,  and $\mathsf{x}\in \mathbb{X}\times \mathbb{L}$.
For the sake of convenience, the following  operators are defined.
\begin{itemize}
	\item $\mathcal{L}(\cdot)$: The projection from $\mathbb{X}\times \mathbb{L}$ to $ \mathbb{L}$, e.g., $\mathcal{L}(\mathsf{x})=\rho$.
	\item $\mathcal{L}^i(\cdot)$: The projection from $\mathbb{X}\times \mathbb{L}$ to $ \mathbb{L}^i$, e.g., , $\mathcal{L}^1(\mathsf{x})=k$.
	\item $\mathcal{L}^{i\times j}(\cdot)$: The projection from $\mathbb{X}\times \mathbb{L}$ to $ \mathbb{L}^i \times \mathbb{L}^j$, e.g.,  $\mathcal{L}^{1 \times 2}(\mathsf{x})=l=(k,i)$,
	$\mathcal{L}^{1 \times 3}(\mathsf{x})=l=(k,g)$.
	\item $\mathcal{X}(\cdot)$: The projection from $\mathbb{X}\times \mathbb{L}$ to $ \mathbb{X}$, e.g., $\mathcal{L}(\mathsf{x})=x$.
	\item $\mathcal{X}\mathcal{L}^{i \times j}(\cdot)$:  The projection from $\mathbb{X}\times \mathbb{L}$ to $ \mathbb{X}\times  \mathbb{L}^i \times \mathbb{L}^j$, e.g., $\mathcal{X}\mathcal{L}^{1 \times 2}(\mathsf{x})=(x,l)$.
\end{itemize}

Thirdly, the  state $x$ is augmented by the above label  $\rho$,
i.e., $\mathsf{x}=(x,\rho)$.
Lastly,  an augmented LRFS is an RFS who has the following form 
\begin{equation}\label{alrf}
	\mathsf{X}=\{\mathsf{x}_1,\mathsf{x}_2,\cdots,\mathsf{x}_n\}.
\end{equation}

For each element $\mathbf{x}=(x,l)$ in the  LMB $\mathbf{X}$, if we subsitute $\mathbf{x}$ by $\mathsf{x}=(x,\rho)$, where $\rho$ is defined in (\ref{alrfsxx}), we will obtain a new LRFS $\mathsf{X}$, which can be called as an agugmented LMB,
and can be defined as follows.

\subsubsection{Agugmented LMB RFS}
An agugmented LMB RFS is an RFS with the following PDF 
\begin{equation}\label{bqqqqq11}
	\pi(\mathsf{X})=\bigtriangleup(\mathsf{X})1_{\alpha(\Psi)}(\mathcal{L}^{1 \times 2}(\mathsf{X}))[\Phi(\mathsf{X};\cdot)]^\Psi
\end{equation}
where  $l=\mathcal{L}^{1 \times 2}(\mathsf{x})$, and
\begin{eqnarray}\label{sxy}
	\Phi(\mathsf{X};\cdot)= \left\{\begin{array}{ll}
		1-r^{(l)}, & \mbox{if $\alpha(l)\notin \mathcal{L}^{1 \times 2}(\mathsf{X})$}, \\
		r^{(l)}p^{(l)}(x), & \mbox{if $(x,\alpha(l))\in \mathcal{X}\mathcal{L}^{1 \times 2}(\mathsf{X})$},
	\end{array} \right.
\end{eqnarray} 
and $\alpha: \Psi \rightarrow \mathbb{L}$ denotes a 1-1 mapping.
If we define 
$\alpha$ as an identity mapping,      (\ref{bqqqqq11}) can be simplified to 
\begin{equation}
	\pi(\mathsf{X})=\bigtriangleup(\mathsf{X})\omega(\mathcal{L}^{1 \times 2}(\mathsf{X}))p^{\mathsf{X}}
\end{equation}
where  
\begin{eqnarray}
	p(x,\rho)&=&p^{(\mathcal{L}^{1 \times 2}(\mathsf{x}))}=p^{(l)}(x), \\
	\omega(L) &=& \prod_{j\in \mathbb{L}^{1 \times 2}}(1-r^{(j)})\prod_{l\in L}\frac{1_{\mathbb{L}^{1 \times 2}}(l)r^{(l)}}{1-r^{(l)}}.
\end{eqnarray}
For simplicity, we  \textcolor{blue}{use the parameter set $\pi=\{(r^{(\rho)},p^{(\rho)})\}^{*}_{\rho \in \mathbb{L}}$} to abbreviate the above PDF.

Similar to $X$ and $\mathbf{X}$, we also call $\mathsf{X}$ as a multi-object state, 
since it also describes the multi-state of multiple objects.
By integrating the augmented label $\rho$ into the multi-object state, one can find that the group attribute information has been incorporated, 
making possible infer the dynamic state, the identity  and group attribute information of individual objects.

\subsection{Statistical Characteristic of  Augmented LRFS}
This subsection will present some preliminary analysis of  the  statistical characteristic of  augmented LRFS.

\begin{proposition}
	An augmented LRFS with state space $\mathbb{X}$ and label space $\mathbb{L}$, i.e., $\mathsf{X}$ denoted in (\ref{alrf}), essentially  is an RFS on the  space $\mathbb{X} \times \mathbb{L}$.
\end{proposition}
\begin{proof}
	Since the individual elements in  $\mathsf{X}$ are random and unordered  variables  on the  space $\mathbb{X} \times \mathbb{L}$,
	and the  number of elements in  $\mathsf{X}$ is a random variable,
	according to the definition of RFS (see Definition 1),  $\mathsf{X}$ is an RFS.
\end{proof}

\begin{proposition}
	An augmented LRFS with state space $\mathbb{X}$ and label space $\mathbb{L}$, i.e., $\mathsf{X}$ denoted in (\ref{alrf}), essentially  is an LRFS on the  space $\mathbb{X} \times \mathbb{L}$.
\end{proposition}
\begin{proof}
	The proof is direct. Since for each element   $\mathsf{x}=(x,\rho)\in \mathsf{X}$, 
	it follows that  $\mathcal{L}^{1\times 2}(\mathsf{x})=l=(k,i)$  is distinct.
	Thus, the label $\rho=(l,g,c)$ also is distinct for each element   $\mathsf{x}=(x,\rho)\in \mathsf{X}$, which is in accordance with the definition  of a labeled RFS (see Definition 2).
\end{proof}

\begin{proposition}
	For an augmented LRFS $\mathsf{X}$,
	its corresponding RFS  $X$ can be obtained by imposing the operator $\mathcal{X}(\cdot)$, i.e.,
	${X}=\mathcal{X}(\mathsf{X})$, its  corresponding LRFS  $\mathbf{X}$ can be obtained by imposing the operator $\mathcal{X}\mathcal{L}^{1 \times 2}(\cdot)$, i.e., $\mathcal{X}\mathcal{L}^{1 \times 2}(\mathsf{X})=\mathbf{X}$.
	An augmented LRFS $\mathsf{X}$, its corresponding RFS  $X$, and its  corresponding LRFS  $\mathbf{X}$ have the same \textcolor{blue}{cardinality} distribution.
\end{proposition}
\begin{proof}
	Since the projection operator does not change the number of elements in $\mathsf{X}$, it is straightforward that the above proposition is true.
\end{proof}

From propositions 1-2, it can be inferred that  the  mathematical operators for both  RFS and  LRFS, are also applicable for the augmented LRFS. Thus, inspired by the  definition  of the   integral operators for both  RFS and  LRFS~\citep{Mahler2007s5a5,Vo2013labeled}, 
the definition of the integral of  an augmented LRFS can be denoted as 
%\begin{eqnarray}
%	&	\int    f(\mathsf{X})\delta \mathsf{X} =&  \nonumber\\
%	&\sum_{i=0}^{\infty}{\frac{1}{i!}  \int_{\rho^i}\int_{\mathbb{X}^i}f(\{(x_1,\rho_1),\cdots,(x_i,\rho_i)\})d(x_1,\cdots,x_i)d\rho_i}.   \nonumber & 
%\end{eqnarray}
\begin{equation*}
	f(\mathsf{X})\delta \mathsf{X} =\sum_{i=0}^{\infty}{\frac{1}{i!}  \int_{\rho^i}\int_{\mathbb{X}^i}f(\{(x_1,\rho_1),\cdots,(x_i,\rho_i)\})d(x_1,\cdots,x_i)d\rho_i}.  
\end{equation*}

For each element $\mathbf{x}=(x,l)$ in the  GLMB $\mathbf{X}$, if we subsitute $\mathbf{x}$ by $\mathsf{x}=(x,\rho)$, where $\rho$ is defined in (\ref{alrfsxx}), we will obtain a new LRFS $\mathsf{X}$, which can be called as an agugmented GLMB.
Thus, an agugmented GLMB can be defined as follows.

\subsubsection{Agugmented GLMB RFS}
An agugmented GLMB  RFS is  an LRFS with the following PDF 
\begin{equation}\label{agugmentedGLMB}
	\pi(\mathsf{X})=\bigtriangleup(\mathsf{X})\sum_{c\in \mathbb{C}}{\omega^{(c)}(\mathcal{L}^{1 \times 2}(\mathsf{X}))[p^{(c)}]^{\mathsf{X}}}
\end{equation}
where   
$\mathbb{C}$ is a discrete index set, and 
\begin{eqnarray}
	\sum_{L \subseteq \mathbb{L}^{1 \times 2}}\sum_{c \subseteq \mathbb{C}}	{\omega(L)}&=&1,\\
	\int p^{(c)}(x,\rho)dx&=&	\int p^{(c)}(x,  \mathcal{L}^{1 \times 2}(x))dx=1.
\end{eqnarray}

\section{System Model}
\subsection{Dynamic Model}

\begin{figure}
	\centering
	\includegraphics[width=0.5\textwidth]{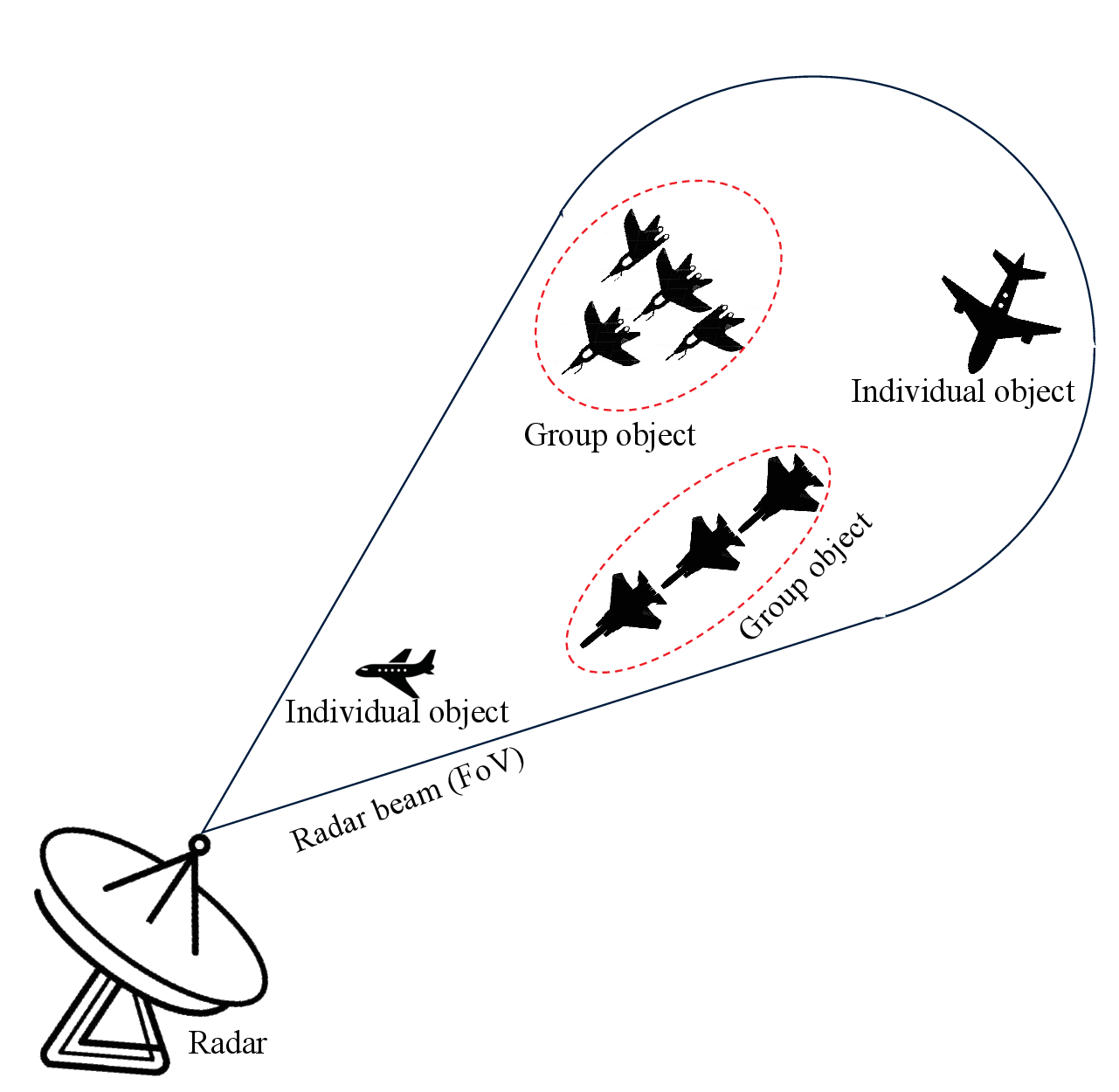}\\
	\caption{{The considered GOT scenario.}}
\end{figure}

The considered GOT scenario is shown in Fig.~2. 
There are multiple objects in the field of view (FoV) of a radar,
some of which are closely positioned and move in a coordinated manner, forming several group objects.
To capture the entire dynamic information of all  objects, it is necessary to model both the dynamics  and group attributes of individual objects.
Here, we use the proposed    augmented LRFS to denote 
the multi-state of these objects  at time $k$, i.e., 
\begin{equation}\label{dynamicrfs}
	\mathsf{X}_{k}=\{\mathsf{x}_{k,1},\mathsf{x}_{k,2},\cdots, \mathsf{x}_{k,n(k)} \}
\end{equation}
where $\mathsf{x}_{k,j}=(x_{k,j},\rho_{k,j})$ denotes the individual  state of the $j$-th object at time $k$,
consisting of dynamic state $x_{k,j}$ and label state $\rho_{k,j}$.
Note that the label state $\rho_{k,j}$ has been defined in (\ref{alrfsxx}), 
and $n(k)$ denotes the number of these object in the FoV  at this time.

Thanks to the introduction of the augmented LRFS, 
it can be found that the dynamics, labels, and group  attributes  of all of these objects at time $k$, are simultaneously contained in the multi-state,  i.e., the augmented LRFS $\mathsf{X}_k$.
Thus, if  $\mathsf{X}_k$ can be accurately estimated, the dynamics, labels, and group  attributes  of individual objects also can be accurately inferred.

%and some of them constitute several group targets
%and there dynamic states at time $k$ can be represented by a random finite set (RFS) ${X}_{k}$,  i.e., 
%\begin{equation}\label{dynamicrfs}
%	X_{k}=\{\boldsymbol{x}_{k,1},\boldsymbol{x}_{k,2},\cdots, \boldsymbol{x}_{k,n(k)} \}
%\end{equation}
%where $\boldsymbol{x}_{k,i}$ denotes the individual dynamic state of the $i$-th target at time $k$, and $n(k)$ denotes the number of the targets in the FoV  at this time.

%Note from Fig.~1 that, some of these targets constitute several group targets since these targets are within a group target are closely spaced and move in a coordinated manner [].

For the dynamic state $x_k \in \mathsf{x}_k$, it is worthing noting that its dynamic state transition function 
is correlated with its corresponding group attribute $\mathcal{L}^{3 \times 4}(\mathsf{x}_k)=(g_k,c_k)$.
On one hand, if $\mathcal{L}^3(\mathsf{x}_k)=g_k=0$, 
which means that the object associated with $\mathsf{x}_k$ dose not belong to any group.
Its dynamics only depend on itself, and is  independent with the dynamics of other objects. 
Thus, we can denote its dynamic state transition function as 
\begin{equation}\label{transtionfunc1}
	f(x_{+}|x,\mathcal{L}^{1 \times 2}(\mathsf{x}))=f(x_{+}|x,l)
\end{equation}
where $x_{+}$ denotes the  dynamic state at the next time, and the time subscript such as 
$k$, has been omitted.
On the other  hand,   
if $\mathcal{L}^3(\mathsf{x}_k)=g_k \neq0$, 
which means that the object associated with $\mathsf{x}_k$ belongs to the group indexed by $g_k$, 
it is reasonable to assume that its dynamic state  not only relies on itself,  but also depends on the dynamic state of this  group.
Thus, it is reasonable to  denote its dynamic state transition function as 
\begin{equation}\label{transtionfunc2}
	f(x_{+}|x,\mathcal{L}^{4}(\mathsf{x}),\mathcal{L}^{1 \times 2}(\mathsf{x}))=f(x_{+}|x,c,l).
\end{equation}

\subsection{Measurement Model}
Suppose that the radar receives $m(k)$ measurements at time $k$, and these measurements  can be also represented by an RFS ${Z}_{k}$,  i.e., 
\begin{equation}\label{measurements}
	Z_{k}=\{\boldsymbol{z}_{k,1},\boldsymbol{z}_{k,2},\cdots, \boldsymbol{z}_{k,m(k)} \}
\end{equation}
where $\boldsymbol{z}_{k,j}$ denotes the  $j$-th measurement at time $k$.
\subsection{The Objective of This Paper}
After formulating  the dynamics of these multiple objects and the measurements, the objectives of this paper are two-fold:
\begin{itemize}
	\item The first is to simultaneously estimate the number of objects and the dynamic states of individual objects, which is in accordance with the objective of  the traditional MOT problem.
	\item The second is to estimate the number and structure of groups present among these objects,  and, at the same time, infer the group attribute for individual objects.
\end{itemize}

Since we have used the augmented LRFS $\mathsf{X}_{k}$ to completely
capture all information of these objects, 
which  includes the dynamic state, identity and group attribute of individual objects.
Then, the objectives of this paper essentially equal to the calculation of the posterior multi-object PDF of $\mathsf{X}_{k}$  conditioned on the measurements at  current time, $Z_k$.
After denoting the  posterior multi-object PDF  as $\pi_k(\mathsf{X}_{k}|Z_{k})$,
we can iteratively calculate it via the following multi-object Bayes filter,
\begin{eqnarray}
	\pi_{k|k-1}(\mathsf{X}_{k}|Z_{k-1})\!&=&\!\!\!\int f_{k|k-1}(\mathsf{X}_{k}|\mathsf{X})\pi_{k-1}(\mathsf{X}|Z_{k-1})\delta \mathsf{X}   \label{bay2}      \label{bay1}\\
	\pi_k(\mathsf{X}_{k}|Z_{k})&=& \frac{g_k(Z_{k}|\mathsf{X}_{k})\pi_{k|k-1}(\mathsf{X}_{k}|Z_{k-1})}
	{\int g_k(Z_{k}|\mathsf{X})\pi_{k|k-1}(\mathsf{X}|Z_{k-1})\delta \mathsf{X}}     \label{bay2}
\end{eqnarray}
where $\pi_{k-1}(\mathsf{X}|Z_{k-1})$  is assumed to given at time $k-1$, 
$	\pi_{k|k-1}(\mathsf{X}_{k}|Z_{k-1})$ represents the predicted multi-object PDF  at time $k-1$,
$f_{k|k-1}(\mathsf{X}_{k}|\mathsf{X})$ and $g_k(Z_{k}|\mathsf{X}_{k})$ denote the multi-object transition kernel function, and the likelihood function at the current time, respectively.

However, before applying the above  multi-object Bayes filter,  two key problems needs to be solved.
The first is  the specific expression of $f_{k|k-1}(\mathsf{X}_{k}|\mathsf{X})$ and $g_k(Z_{k}|\mathsf{X}_{k})$,
and the second is the feasible solution to the above multi-object Bayes filter  due to 
the involvement of intricate integral calculations.

\section{ Augmented LRFS-based Filter for  GOT Problem}
In  this section, after presenting the specific expression of $f_{k|k-1}(\mathsf{X}_{k}|\mathsf{X})$ and $g_k(Z_{k}|\mathsf{X}_{k})$  in (\ref{bay1})-(\ref{bay2}),
a feasible solution to the above multi-object Bayes filter will be provided, 
which achieves the objectives of this paper.

\subsection{Multi-object Likelihood Function}

For each state $\mathsf{x}$ in a given multi-state $\mathsf{X}$, it 
has two possible choices. First,
it may be  detected with probability $p_D(\mathsf{x})$
and generates a measurement $z$ with likelihood function $g(z|\mathsf{x})$.
Second, it may be not detected with probability $1-p_D(\mathsf{x})$.
Thus, the generated measurement can be formulated by a Bernoulli RFS with parameters $p_D(\mathsf{x})$ and $g(\cdot|\mathsf{x})$.
Furthermore, if  all states in  $\mathsf{X}$ are mutually independent, it can be inferred that all measurements generated by $\mathsf{X}$ form 
a multi-Bernoulli RFS $W$  distributed according to 
\begin{equation}
	\pi_D(W|\mathsf{X})=\{p_D(\mathsf{x}),g(\cdot|\mathsf{x})\}_{\mathsf{x}\in \mathsf{X}}.
\end{equation}

In addition to measurements originating from $\mathsf{X}$, 
$Z$ also includes measurements from clutters denoted as  $C$, i.e., $Z=W\cup C$,
and
$C$ is modeled by a Possion RFS distributed~\citep{Reuter2014label} as
\begin{equation}
	\pi_K(C)=e^{-<c,1>}\kappa^C,
\end{equation}
where $\kappa(\cdot)$ is  the intensity function.
Then, according to FISST~\citep{Mahler2007s5a5,Vo2013labeled}, the multi-object likelihood function  can be calculated as 
\begin{equation}\label{likehihoodfunction}
	g_k(Z_{k}|\mathsf{X}_{k})=\sum_{W \subseteq Z_k} {\pi_D(W|\mathsf{X}_k)\pi_K(Z_k-W)}
\end{equation}
which is exactly the equation (18) in Ref.~\cite{Vo2013labeled}, regardless of the difference between an augmented LRFS  and an LRFS.

\subsection{Multi-object  Transition Kernel}
For the object $\mathsf{x}\in \mathsf{X}$, it is identified by a unique label $\mathcal{L}^{1 \times 2}(\mathsf{x})=l=(k,i)$.
As mentioned in Proposition 2,   $l=(k,j)$ is unique and distinct.
It is straightforward that $\mathcal{L}^{1 \times 2}(\mathsf{x})=l=(k,i)$  wil be perserved during the surviving time of this object.
In other words,
for each state $\mathsf{x}=(x,\rho)=(x,l,g,c)$, if it continues to exist at the next time  with probability  $p_S(x,l)$ and move to a new state  $\mathsf{x}_{+}=(x_{+},\rho_{+})=(x_{+},l_{+},g_{+},c_{+})$,
we have $\delta_l(l_{+})=1$.

Furthermore, if  it continues to exist at the next time  with probability  $p_S(x,l)$ and move to a new state,
depending on whether the object belongs to a group or not, its transition function has different forms.
Specifically, according to section 3.1, 
if it does not belong to any group at current time,  its transition function will be $f(x_{+}|x,l)\delta_l(l_{+})$, i.e., equation (\ref{transtionfunc1}).
Inversely,
if it belongs to a group with group label $g$ at the current time, it should follow the  transition function  $f(x_{+}|\mathcal{L}^4(\mathsf{x}),l)\delta_l(l_{+})=f(x_{+}|c,l)\delta_l(l_{+})$, i.e., equation (\ref{transtionfunc2}).
In summary, for the object with current state $\mathsf{x}=(x,\rho)=(x,l,g,c)$, 
if it continues  to exist at the next time  with probability  $p_S(x,l)$ and moves to a new state  $\mathsf{x}_{+}=(x_{+},\rho_{+})=(x_{+},l_{+},g_{+},c_{+})$,
the transition PDF can be represented as 
%\begin{eqnarray}  \label{transccccc}
%	&	f(\mathsf{x}_{+}|\mathsf{x})=&  \nonumber\\
%	&f(x_{+}|x,l)\delta_l(l_{+})\delta_0(g)+f(x_{+}|c,l)\delta_l(l_{+})(1-\delta_0(g)).  & 
%\end{eqnarray}
\begin{equation} \label{transccccc}
		f(\mathsf{x}_{+}|\mathsf{x})=f(x_{+}|x,l)\delta_l(l_{+})\delta_0(g)+f(x_{+}|c,l)\delta_l(l_{+})(1-\delta_0(g)). 
\end{equation}
On the other hand, except surviving,   the state $\mathsf{x}=(x,\rho)=(x,l,g,c)$ also has the  probability $q_S(x,l)=1-p_S(x,l)$ to die at the next time.
Therefore,  the set of the state  $\mathsf{x}_{+}$ is a  Bernoulli RFS with argumented label  $\rho_{+}$.

On premise of that  the state transition of all  objects in the current multi-state $\mathsf{X}$ are multually independent, 
it follows that the set of surviving objects at the next time is an augmented LMB RFS 
with parameter set
$\{p_S(x,l), f(\cdot|\mathsf{x})\}^{*}_{\mathsf{x}\in \mathsf{X}}$, denoted by $\mathsf{W}$.
Hence,  according to (\ref{bqqqqq11}),  $\mathsf{W}$ is subjected to the distribution  
\begin{equation}
	f_S(\mathsf{W}|\mathsf{X})=\Delta(\mathsf{W})\Delta(\mathsf{X})1_{\mathcal{L}^{1 \times 2}(\mathsf{X})}(\mathcal{L}^{1 \times 2}(\mathsf{W})
	[\Phi(\mathsf{W};\cdot)]^{\mathsf{X}}
\end{equation}
where
%\begin{small}
%\begin{equation}
%\Phi(\mathsf{W};\mathsf{X})=\sum_{\mathsf{x} \in \mathsf{W}}{p_S(x,l)
	%f(\mathsf{x}_{+}|\mathsf{x})\delta_l(l_{+})+[1-1_{\mathcal{L}^{1 \times 2}(\mathsf{W})}(l)]q_S(x,l)}.
%\end{equation}
%\end{small}
\begin{eqnarray}\label{sx11y}
	\Phi(\mathsf{W};\mathsf{X})= \left\{\begin{array}{ll}
		q_S(x,l), & \mbox{if $l\notin {\mathcal{L}^{1 \times 2}(\mathsf{W})}$}, \\
		p_S(x,l)f(\mathsf{x}_{+}|\mathsf{x}), & \mbox{if $(x,l)\in \mathsf{W}$},
	\end{array} \right.
\end{eqnarray}
If  we take the augmented LRFS $\mathsf{W}$ as a special LRFS, one can find that the above equation is exactly the equation (25) in Ref.~\cite{Vo2013labeled} .

Apart from the surviving objects, at the next time, there maybe exist the newborn objects.
According to Ref.~\cite{Reuter2014label}, the set of newborn objects at the next time  can be  formulated as an LMB RFS  $\mathbf{Y}$ distributed according to  
\begin{equation}
	f_B(\mathbf{Y})=\Delta(\mathbf{Y})w_B(\mathcal{L}(\mathbf{Y}))[p_B]^\mathbf{Y}
\end{equation}
where $p_B$ is the newborn probability.
However, the above formulation does not consider the group information of the newborn objects.
Thus, for $\mathbf{x}=(x,l)\in \mathbf{Y}$, we extend $\mathbf{x}$ to $\mathsf{x}=(x,\rho)=(x,l,g,c)$, and then we obtain a new augmented LRFS for newborn objects, i.e.,  $\mathsf{Y}$, that follows the PDF  
\begin{equation}
	f_B(\mathsf{Y})=\bigtriangleup(\mathsf{Y})\omega(\mathcal{L}^{1 \times 2}(\mathsf{Y}))p^{\mathsf{Y}}.
\end{equation}
Assume that each newborn object is  independent and does not belong to any group at the borntime, then for 
any $\mathsf{x} \in \mathsf{Y}$, we set $\mathcal{L}^3(\mathsf{x})=g=0$.

It follows that the multi-object state at the next time, $\mathsf{X}_{+}$,  is the combination of newborn and surviving objects, i.e.,  $\mathsf{X}_{+}=\mathsf{W} \cup \mathsf{Y}$.
Then, according to FISST~\citep{Mahler2007s5a5,Vo2013labeled},   the multi-object transition kernel is
\begin{equation}\label{transitionkernel}
	f(\mathsf{X}_{+}|\mathsf{X})=\sum_{\mathsf{W}\subseteq \mathsf{X}}
	{f_S(\mathsf{W} |\mathsf{X})f_B(\mathsf{X}_{+}-\mathsf{W} )}.
\end{equation}
which is exactly equation (31)  in Ref.~\cite{Vo2013labeled}, regardless of the difference between an augmented LRFS  and an LRFS.

\begin{lemma}
	If the current multi-object prior  $\pi_{k-1}(\mathsf{X}|Z_{k-1})$ in (\ref{bay1})  is an agumented GLMB  RFS of the form (\ref{agugmentedGLMB}),
	then both the predicted and posterior  multi-object  PDF are aslo agumented GLMB RFSs  of the form (\ref{agugmentedGLMB}).
	In other words, 
	the augmented GLMB RFS is closed under the multi-object  Bayes filter (\ref{bay1})-(\ref{bay2})
	with respect to the multi-object likelihood function (\ref{likehihoodfunction}) and multi-object transition kernel (\ref{transitionkernel}).
	Moreover,  the multi-object Bayes filter can be analytically
	solved by the GLMB filter proposed in Ref.~\cite{Vo2013labeled}, and further can be approximatively solved by the  LMB filter proposed in Ref.~\cite{Reuter2014label}.
\end{lemma}
\begin{proof}
	The proof can be explained from two aspects.
	First, an agumented GLMB  RFS still is essentially a GLMB RFS, we can consider  it as a special GLMB RFS with agumented label space. 
	Moreover, based on this consideration,
	the multi-object likelihood function (\ref{likehihoodfunction}) and multi-object transition kernel (\ref{transitionkernel}) are equivalent to equation (18) and (31) in Ref.~\cite{Vo2013labeled}.
	Second, according to Ref.~\cite{Vo2013labeled}, 	the GLMB RFS is closed under the multi-object  Bayes filter (\ref{bay1})-(\ref{bay2})
	with respect to the multi-object likelihood function (i.e.,  equation (18) in Ref.~\cite{Vo2013labeled}) and multi-object transition kernel (i.e.,  equation (31) in Ref.~\cite{Vo2013labeled}), and the multi-object Bayes filter can be analytic solved by the GLMB filter. Moreover, from Ref.~\cite{Reuter2014label}, the multi-object  Bayes filter can be approximatively solved by the  LMB filter.
	Thus, we can infer that this lemma holds true.
\end{proof}

\subsection{Augmented LRFS-based Filter}
According to Lemma 1, after formulating the GOT problem via the
agumented LRFS,
it  can be approximatively solved by the  LMB filter.
Thus, in this subsection, 
we will develop an augmented LRFS-based filter, 
which combines the LMB filter with the agumented LRFS,  simultaneously integrating group information into the LMB filter.
Differing from the original LMB filter~\citep{Reuter2014label} that includes the prediction and update steps, the new LMB filter consists of three steps:  state prediction, state update and group information update.

%integrates the group information from two aspects.
%First, at the prediction part, the prediction of each LMB component is different.
%If the LMB component  represents the target does not belong to any group, then the transition fuction $f(x_{+}|x,l)$ is used, otherwise, the transition fuction $f(x_{+}|c,l)$ is used.
%Second, After update step,  we add a new step  named the extraction of group information to update the group information of all targets.

\subsubsection{State Prediction}
First, let's begin with the state prediction step of the new LMB filter, whch can be presented by the following proposition.
\begin{proposition}
\textcolor{blue}{If the following premises are true,
	\begin{itemize}
		\item the posterior multi-object  PDF has  the parameter set $\pi=\{r^{(\rho)},p^{(\rho)}\}^{*}_{\rho \in \mathbb{L}}$ with state space $\mathbb{X}$ and  label space $\mathbb{L}$, 
		\item  the multi-object birth model has  the parameter set  	$\pi=\{r^{(\rho)}_{B},p^{(\rho)}_{B}\}^{*}_{\rho \in \mathbb{B}}$, with 
		state space $\mathbb{X}$ and label space $\mathbb{B}$,
	\end{itemize}
	then,  the predicted multi-object   PDF has the parameter set 
		\begin{equation}
		\pi_{+}=\{(r_{+,S}^{(\rho)},p_{+,S}^{(\rho)})\}^{*}_{\rho\in \mathbb{L}}\cup \{(r_{B}^{(\rho)},p_{B}^{(\rho)})\}^{*}_{\rho \in \mathbb{B}},
	\end{equation}
	where
	\begin{eqnarray}
		\rho&=&(l,g,c), \\
		r_{+,S}^{(\rho)},&=& \eta_S(\rho)r^{(\rho)}, \\
		\eta_S(\rho)&=&<p_S(\cdot,\rho),p(\cdot,\rho)>,\\
		p_{+,S}^{(\rho)}&=& <p_S(\cdot,\rho)f(\mathsf{x}_{+}|\mathsf{x}),p(\cdot,\rho)>/ \eta_S(\rho),
	\end{eqnarray}
	and $f(\mathsf{x}_{+}|\mathsf{x})$ is defined in (\ref{transccccc}).}
	%and label space $\mathbb{L}_{+}=\mathbb{B}\cup \mathbb{L}$.
	
%	Suppose that the multi-target posterior PDF is an agumented LMB RFS with state space $\mathbb{X}$, label space $\mathbb{L}$ and parameter set $\pi=\{r^{(\rho)},p^{(\rho)}\}_{\rho \in \mathbb{L}}$,
%	and the multi-target birth model is an agumented LMB RFS with state space $\mathbb{X}$, label space $\mathbb{B}$ and parameter set
%	$\pi=\{r^{(\rho)}_{B},p^{(\rho)}_{B}\}_{\rho \in \mathbb{B}}$,
%	then the multi-target predicted  PDF is also an agumented LMB RFS with state space  $\mathbb{X}$, label space $\mathbb{L}_{+}=\mathbb{B}\cup \mathbb{L}$,
%	which has the parameter set 
%	\begin{equation}
%		\pi_{+}=\{(r_{+,S}^{(\rho)},p_{+,S}^{(\rho)}\}_{\rho\in \mathbb{L}}\cup \{(r_{B}^{(\rho)},p_{B}^{(\rho)})\}_{\rho \in \mathbb{B}},
%	\end{equation}
%	where
%	\begin{eqnarray}
%		\rho&=&(l,g,c), \\
%		r_{+,S}^{(\rho)},&=& \eta_S(\rho)r^{(\rho)}, \\
%		\eta_S(\rho)&=&<p_S(\cdot,l),p(\cdot,\rho)>,\\
%		p_{+,S}^{(\rho)}&=& <p_S(\cdot,l)f(\mathsf{x}_{+}|\mathsf{x}),p(\cdot,\rho)>/ \eta_S(\rho),
%	\end{eqnarray}
%	and $f(\mathsf{x}_{+}|\mathsf{x})$ is defined in (\ref{transccccc}).
\end{proposition}

\subsubsection{State Update}
The state update step of the new LMB filter can be presented by the following proposition.

\begin{proposition}
\textcolor{blue}{If the predicted multi-object   PDF has the parameter set $\pi_{+}=\{r^{(\rho)}_{+},p^{(\rho)}_{+}\}^{*}_{\rho \in \mathbb{L}_{+}}$
	with state space $\mathbb{X}$ and label space $\mathbb{L}_{+}$,
	then the posterior multi-object  PDF can be updated by   the parameter set 
	$\pi(\mathsf{X}|Z)=\{r^{(\rho)},p^{(\rho)}(\cdot)\}^{*}_{\rho\in \mathbb{L}_{+}}$
	where}
	\begin{equation}
		p^{(\rho)}\!\! = \frac{1}{r^{(\rho)}}  \!\!\! \!\!\! \sum_{(I_{+},\theta)\in \mathcal{F}(\mathbb{L}_{+})\times \Theta_{I_{+}}}{\omega^{(I_{+},\theta)}(Z)1_{I_{+}}(\rho)p^{(\theta)}(x,\rho)},     
	\end{equation}
	\begin{eqnarray}
		r^{(\rho)} \!\!\!  &=&  \!\!\!  \sum_{(I_{+},\theta)\in \mathcal{F}(\mathbb{L}_{+})\times \Theta_{I_{+}}}{\omega^{(I_{+},\theta)}(Z)1_{I_{+}}(\rho)},\\
		\omega^{(I_{+},\theta)}(Z)&\varpropto&\omega_{+}(I_{+})[\eta^{(\theta)}_Z]^{I_{+}},\\
		p^{(\theta)}(x,\rho|Z)&=&\frac{p_{+}(x,\rho)\psi_Z(x,\rho;\theta)}{\eta_Z^{(\theta)}(\rho)},\\
		\eta_Z^{(\theta)}(\rho)&=&<p_{+}(\cdot,\rho),\psi_Z(\cdot,\rho;\theta)>, \\
		\omega_{+}(I_{+}) &=& \prod_{j_{+}\in \mathbb{L}_{+}}(1-r_{+}^{(j)})\prod_{\rho\in L_{+}}\frac{1_{\mathbb{L}_{+}}(\rho)r_{+}^{(\rho)}}{1-r_{+}^{(\rho)}}.
	\end{eqnarray}
	
	\begin{eqnarray}
		\psi_Z(x,\rho;\theta)= \left\{\begin{array}{ll}
			\frac{p_D(x)g(z_{\theta(\rho)}|x,l)}{\kappa(z_{\theta(\rho)})}, & \mbox{if $\theta(\rho)>0$}, \\
			1-p_D(x), & \mbox{if $\theta(\rho)=0$},
		\end{array} \right.
	\end{eqnarray}
	and $\Theta_{I_{+}}$ is the space of unique mappings  from $\theta$: $I_{+}$ to $ \{0,1,\cdots,|Z|\}$,
	$\kappa(\cdot)$ is defined in (21), 
	and $g(z|x,\rho)$ is the single object likelihood.
\end{proposition} 

The proof of propositions 4-5 can be explained as follows.
The two propostions with LRFS version, i.e., 	propostions 2 and 4 in Ref.~\cite{Reuter2014label},
have been proved in Ref.~\cite{Reuter2014label}.
Since these propostions can be taken as  the propostions 2-4 in Ref.~\cite{Reuter2014label} with agumented LRFSs,
and an agumented LRFS is essentially an LRFS,
thus, it follows that propositions 4-5 hold true.

After the state of state update, the number of objects $\hat{n}_k$ can be extracted by  calculating the maximum a posteriori (MAP) estimation of the  cardinality of 
the estimated agumented LMB  RFS $\mathsf{X}$~\citep{Yang2023label,Vo2013labeled}.
Then, we pick up $\hat{n}_k$ Bernoulli items with the
highest weight,  calculate the corresponding mean  via $\hat{x}^{({\rho})}=\int xp^{(\rho)}(x)dx$ for each Bernoulli item  $(r^{(\rho)},p^{(\rho)})$, 
and we take this mean as  the estimated  dynamic state of the potential object represented by this  Bernoulli item.
Finally, we obtain the estimation of the dynamic states of individual objects,
which achieves the first objective of this paper.

\subsubsection{Group Information Update}
After obtaining the approximated multi-object posterior
PDF   $\pi(\mathsf{X}|Z)=\{r^{(\rho)},p^{(\rho)}(\cdot)\}^{*}_{\rho\in \mathbb{L}_{+}}$,
we can further extract the group information of each Bernoulli item $(r^{(\rho)},p^{(\rho)})$.
Note that if the number of Bernoulli item in $\pi(\mathsf{X}|Z)$ equals to one, 
which means that there exist at most only one object at this time,  then the update of group information can be skipped,
since one object can not form a group.

If the number of   the Bernoulli items  in $\pi(\mathsf{X}|Z)$ is $n$, where $n>1$, 
first, for each Bernoulli item  $(r^{(\rho)},p^{(\rho)})$,    we calculate its corresponding mean via $\hat{x}^{({\rho})}=\int xp^{(\rho)}(x)dx$, which can be taken as the estimated  dynamic state of the potential object represented by this  Bernoulli item.
Second,  for all Bernoulli items in $\pi(\mathsf{X}|Z)$, we calculate the adjacent matrix given by 
\begin{equation}\label{direct}
	A_{k}=\left[ \begin{array}{cccc}
		0    & a_k(1,2)    &\cdots    &a_k(1,n)\\
		a_k(2,2) &0        &\cdots   &a_k(2,n) \\
		\vdots     &\vdots   & \ddots  & \vdots \\
		a_k(n,1)   &a_k(n,2)   &\cdots   &0
	\end{array} \right]
\end{equation}
where 
\begin{eqnarray}\label{proof333}
	a_k({i,j})= \left\{\begin{array}{ll}
		1, & \mbox{if $i\neq j$ and $d_k({i,j})\leq \varepsilon$}, \\
		0, & \mbox{otherwise},
	\end{array} \right.
\end{eqnarray}
and $d_k({i,j})=||\hat{x}^{i}-\hat{x}^{j}||_2$  and $ \varepsilon$  is a given threshold.
According to the adjacent matrix $A_k$, we can construct a graph $G_k$.

Then, we calculate  all the connected components\footnote{The definition of connected component: A connected component is a set of vertices within a graph such that there is a path between every pair of vertices in the set, and no path exists to connect any vertex in the set to a vertex outside the set~\citep{Trudeau2003intro}.} of the graph $G_k$.
We take each connected component as a group, where all Bernoulli items belonging to this connected component are regarded as the members of this group. 
Furthermore,  we interpret the graph structure of this connected component as the group's structure.
For this group, we assign a new distinct integer $g'$  as its index, and take  the 
spatial centroid of this group as its \textcolor{blue}{geometry} center  $c'$.

Lastly, for each Bernoulli item   $(r^{(\rho)},p^{(\rho)})$,  we update its group information and integrate them into its augmented label $\rho=(l,g,c)$.
Specifically,    
$g$ is updated by $g'$, which is  the index of the group  that  the Bernoulli item belongs to.
$c$ is  updated by $c'$,  which is the \textcolor{blue}{geometry} center of the group  that  the Bernoulli item belongs to.

In summary,  the entire schematic of the step of group information update is shown in Fig.~3.
For illustration, we take Fig.~4 for example.
Suppose that there are six Bernoulli items after the state update step of the proposed LMB filter with augmented LRFS, which represent the six objects in Fig.~4.
First, we calclulate the adjacent matrix according to (\ref{direct}), yielding 
\begin{equation}\label{direct1111}
	A_{k}=\left[ \begin{array}{cccccc}
		0    &1   &1   &0 &0 &0\\
		1 &0        &1   &0 &0 &0\\
		1    &1   &0 & 0 &0 &0 \\
		0  &0   &0   &0 &1  &0\\
		0 &0 &0 &1 &0 &0\\
		0 &0 &0 &0 &0 &0
	\end{array} \right].
\end{equation}
Then, we calculate  all the connected components.
As shown in Fig.~4, three connected components are calculated, which divides  these objects into three groups.
Lastly, we assign the group  index $g$ and the group \textcolor{blue}{geometry} center  $c$ to each group,
update the group information and integrate them into its augmented label $\rho=(l,g,c)$.

\begin{figure}
	\centering
	\includegraphics[width=0.5\textwidth]{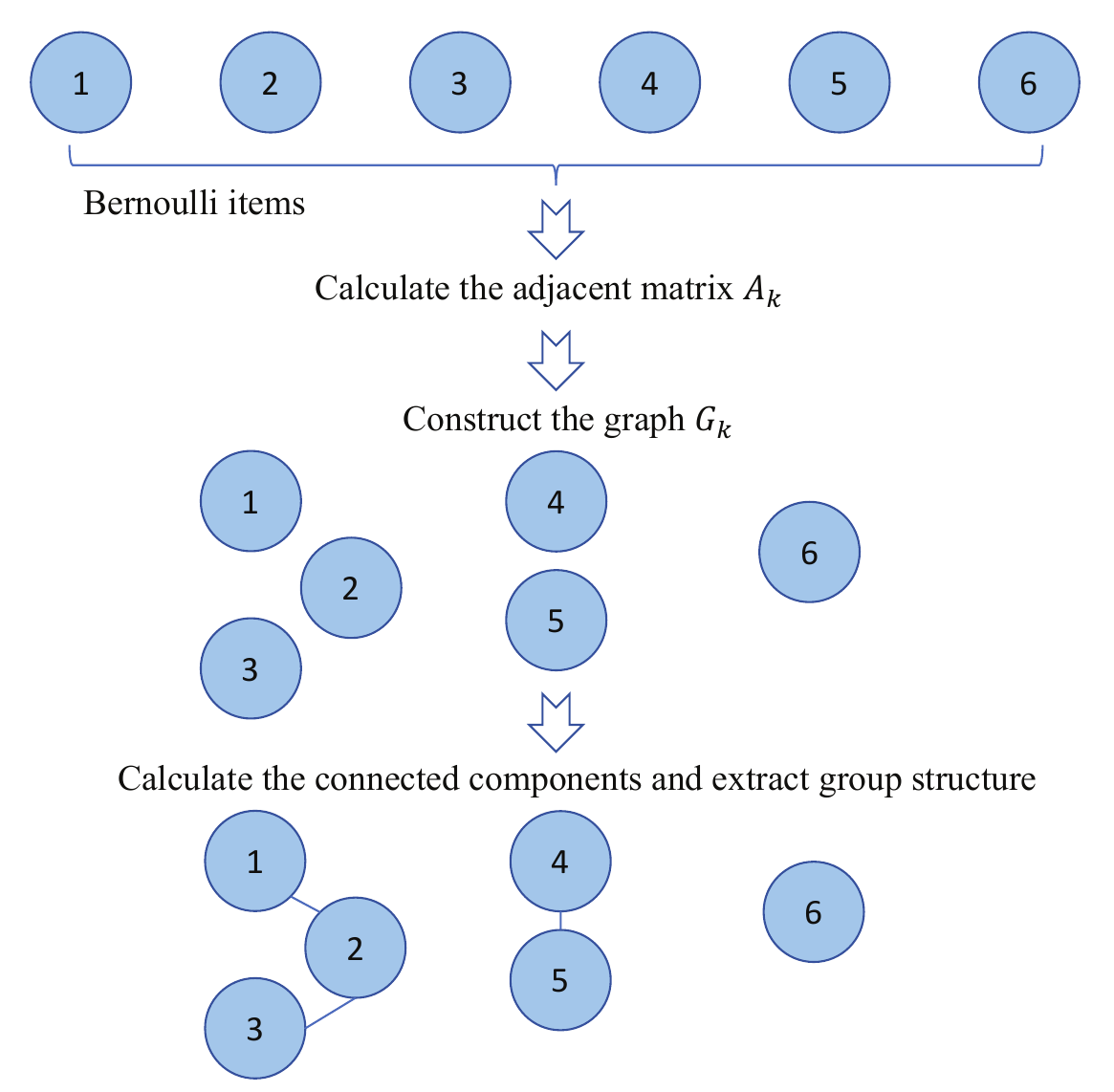}\\
	\caption{{The entire schematic of the step of group information update.}}
\end{figure}

\begin{figure}
	\centering
	\includegraphics[width=0.5\textwidth]{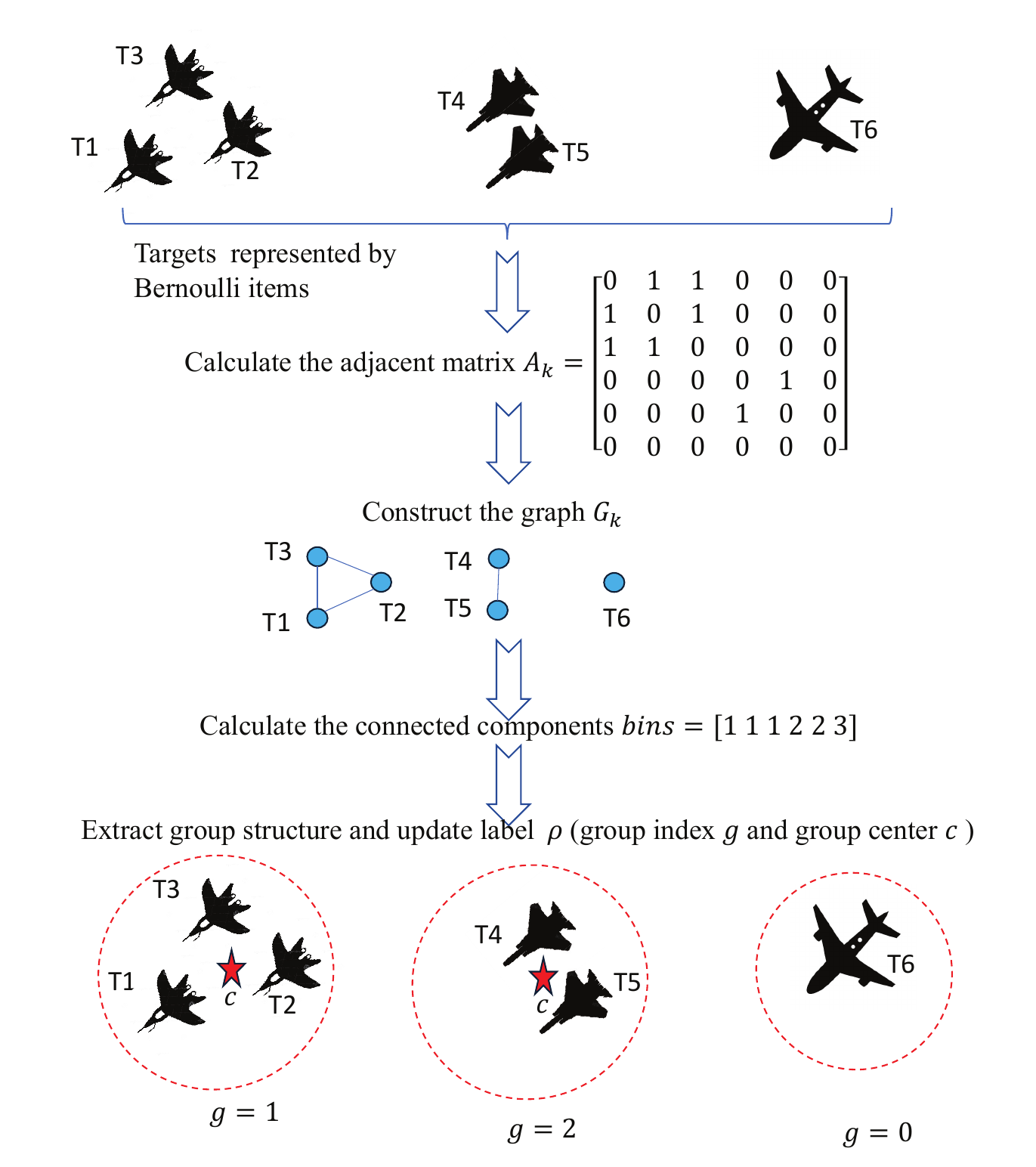}\\
	\caption{{An example for group information extraction, where T$i$ denotes the $i$-th object.}}
\end{figure}

After the state of group information update,
The number of groups and the group structure can be extracted from 
$\mathcal{L}(\mathsf{X})=\{\rho_1,\rho_2,\cdots,\rho_{n(k)}\}$.
First, the number of groups equals to the number of non-zero elements in $\mathcal{L}^3(\mathsf{X})$.
Then, we can find all the same and non-zero elements in $\mathcal{L}^3(\mathsf{X})$, and take them as a group, and the corresponding group \textcolor{blue}{geometry} center can be extracted from $\mathcal{L}^4(\mathsf{X})$.

\section{Performance Evaluation} 
\subsection{Parameters Settings} 
Consider  a two-dimensional scenario with up to 6 objects.
For each object, its state is described by the vector ${x}_k=[p_{x,k},\dot{p}_{x,k},p_{y,k},\dot{p}_{y,k}]^T$,
where $(p_{x,k},p_{y,k})$ and $(\dot{p}_{x,k},\dot{p}_{y,k})$ are the position and velocity of this object at time $k$, respectively.
On the one hand, for the object (with label $l$) that does not belong to any group,
its transition density is given by $f(x_{+}|x,l)=N(x_{+};Fx,Q)$,
where 
\begin{equation}
	F=\left[ \begin{array}{cccc}
		0    &\bigtriangleup t    &0   &0\\
		0     &1       &0   &0 \\
		0     &0 &1  &\bigtriangleup t   \\
		0   &0   &0   &1
	\end{array} \right]
\end{equation}
$\bigtriangleup t$ is time interval and $Q=\varrho^2GG^T$,
$\varrho=5$,
and
\begin{equation}
	G=\left[ \begin{array}{cc}
		(\bigtriangleup t ) ^2/2      &0\\
		\bigtriangleup t        &0 \\
		0     &(\bigtriangleup t ) ^2/2  \\
		0   &\bigtriangleup t  
	\end{array} \right]
\end{equation}
On the other hand, for the object (with label $l$) that belongs to a group with \textcolor{blue}{geometry} center $c$,
its transition density is denoted by $f(x_{+}|x,c,l)$.
Suppose the transition density of the group \textcolor{blue}{geometry} center follows $f(c_{+}|c)=N(c_{+};Fc,Q)$,
then  we have $f(x_{+}|x,c,l)=N(x_{+};(F-I)c+x,Q)$ \footnote{The proof can be found in Appendix.}.
% Track Pruning and Extraction
The survival probability of the objects is set to be $p_S=0.99$.
The birth LMB RFS is set to be $\pi_B=\{(r_B^{(i)},p_B^{(i)})\}_{i=1}^6$,
where $r_B^{(i)}=0.03$,
$p_B^{(i)}=N(x;m_B^{(i)},P_B)$,
$m_B^{(1)}=[-800,0,600,0]^T$,  $m_B^{(2)}=[-800,0,-200,0]^T$, $m_B^{(3)}=[-850,0,-200,0]^T$,
$m_B^{(4)}=[-750,0,-200,0]^T$,  $m_B^{(2)}=[-650,0,670,0]^T$, $m_B^{(3)}=[-750,0,530,0]^T$,
$P_B=10^2I_{4\times 4}.$

The radar receives range measurements over the region $[-1000,1000]m \times [-1000,1000]m$.
The standard deviations of the measurement noise is set to be $\varrho_r=10$.
The detection probability of this radar is set to be $p_D=0.98$.
The clutter measurements follow the Possion RFS with $\kappa(z)=\lambda U(z)/V$,
where $\lambda=30$,
$U(\cdot)$ represents a uniform density over the observed region and $V=\int U(z)dz$.

The performance of the proposed LMB filter with agumented LRFS is compared with the traditional  LMB filter~\citep{Reuter2014label}.
In both the two filters, the parameters such as the maximal number of  components and the pruning threshold
are the same.
In addition, for the proposed LMB filter with agumented LRFS, 
the group threshold $\varepsilon$ in (\ref{proof333}) is set to be $100m$.

\subsection{Simulation Results} 
The   trajectories of all objects are presented in Fig.~5.
One can find that these objects can be divided into two groups, where objects with label 1, 2, and 3 form  group 1, and the rest belongs to  group 2.

\begin{figure}
	\centering
	\includegraphics[width=0.5\textwidth]{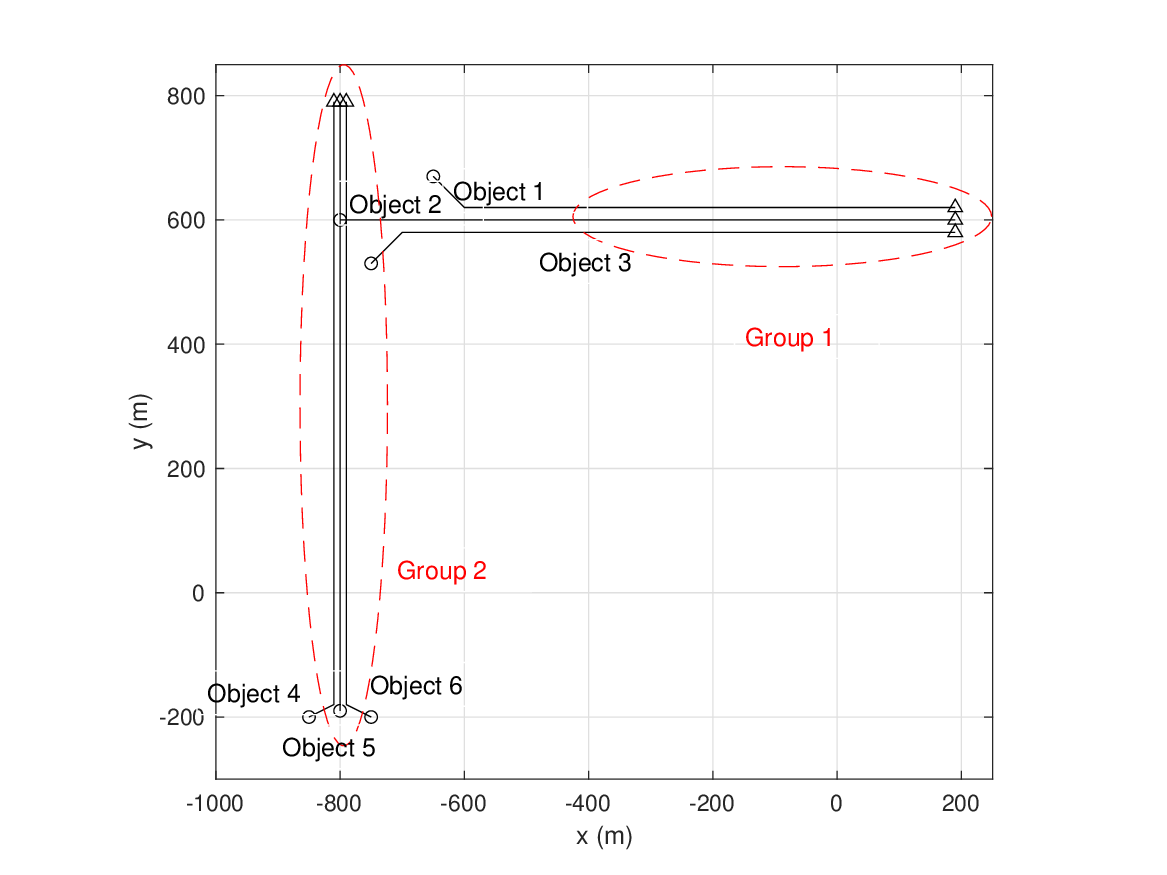}\\
	\caption{{The   trajectories of objects.}}
\end{figure}

Fig.~6 and Fig.~7 shows the tracking results of the proposed LMB filter with the agumented LRFS and the LMB filter over one trail, respectively. 
It can be seen that both the two filters can successfully tracks all of the objects.
Fig.~8 shows the corresponding cardinality estimates of the two filters.
Obviously, both the two filters can accurately estimate the number of objects in most time.

However, due to the  integration of group information, the proposed LMB filter with agumented LRFSs 
is expected to perform better performance than the LMB filter,
which can be verified by Fig.~9.
Fig.~9 illustrates the corresponding average OSPA distance~\citep{4567674,Chen1,Shi2} of the two filters over 100 Monte Carlo (MC) trials. 
It can be seen that the proposed LMB filter with agumented LRFS performs  better  than the LMB filter.
In other words,  comparing with the LMB filter, 
the  proposed LMB filter with agumented LRFSs achieves smaller OSPA distance, which means better tracking 
performance.

Lastly, we investigate the estimated results of group number.
As shown in Fig.~10, the proposed filter with the agumented LRFS can accurately estimate the number of groups in most time.
As a comparison, the LMB  filter does not have the ability to estimate the number of groups due to the lack of the intergation of group  information.

\begin{figure}
	\centering
	\includegraphics[width=0.5\textwidth]{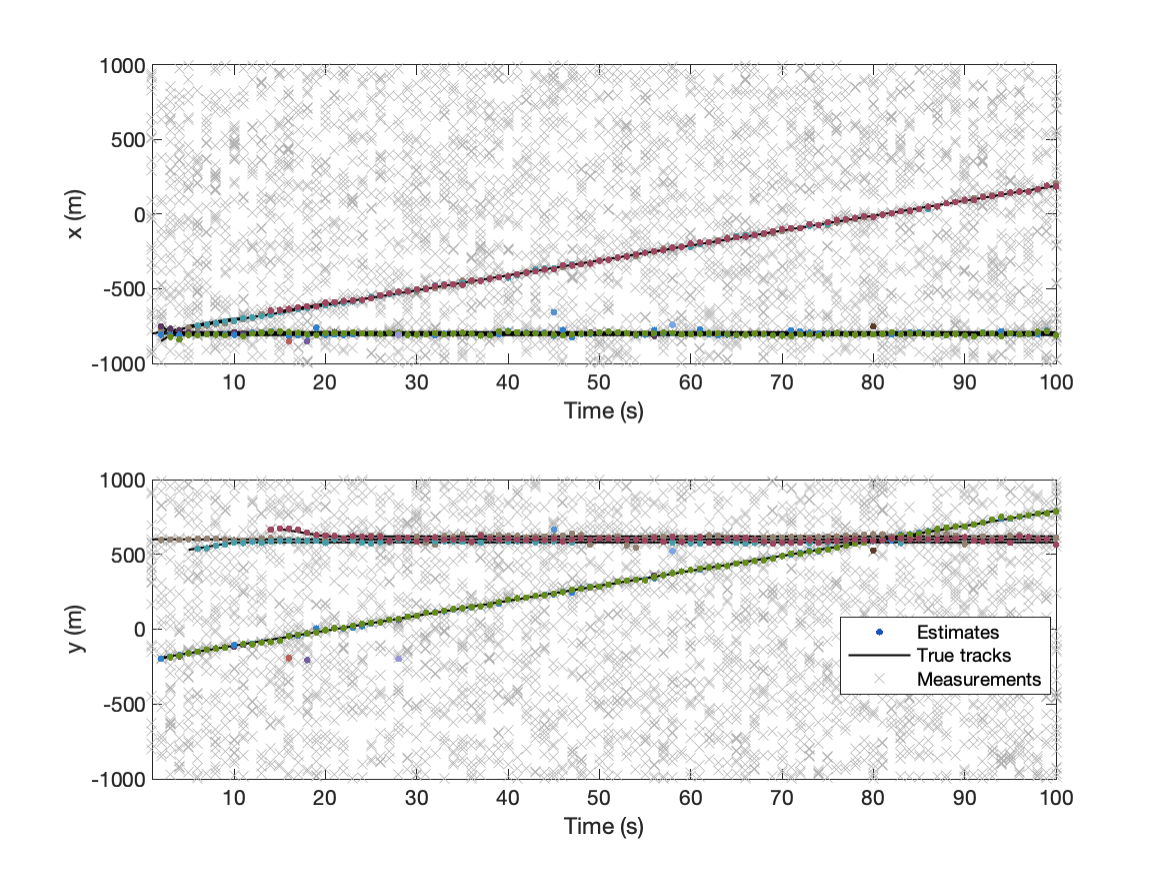}\\
	\caption{{Measurements, true trajectories, and the tracking results of the proposed LMB filter with agumented LRFS.}}
\end{figure}

\begin{figure}
	\centering
	\includegraphics[width=0.5\textwidth]{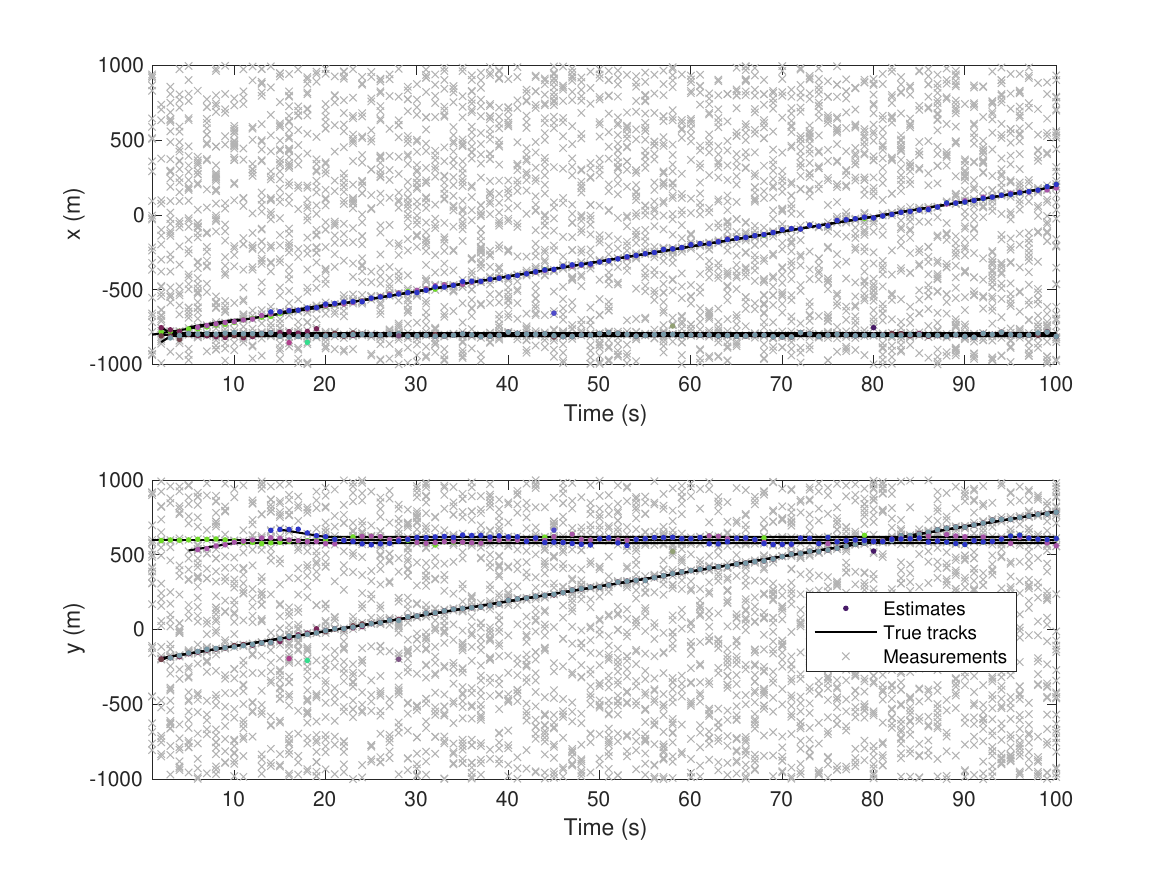}\\
	\caption{{Measurements, true trajectories, and the tracking results of the  LMB filter.}}
\end{figure}

\begin{figure}
	\centering
	\includegraphics[width=0.5\textwidth]{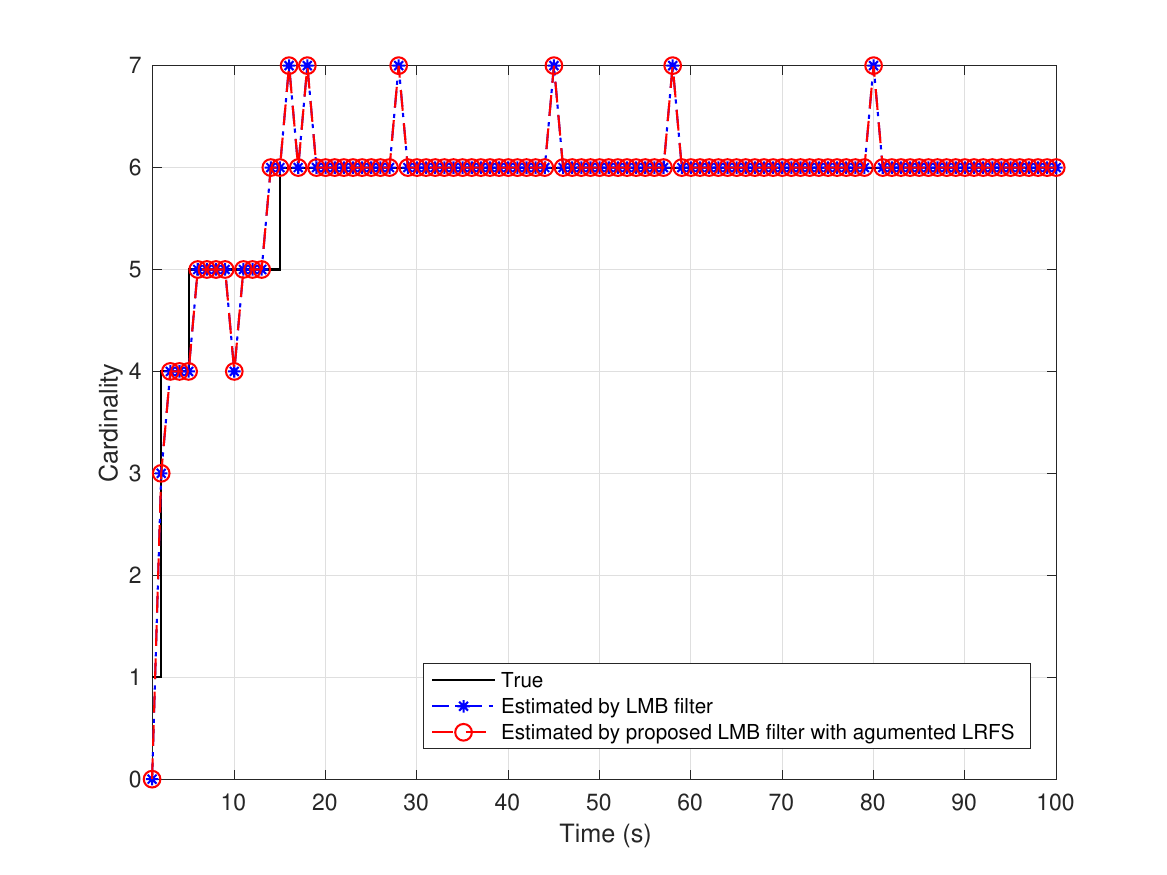}\\
	\caption{{Cardinality estimates.}}
\end{figure}

%\begin{figure}
%	\centering
%	% Requires \usepackage{graphicx}
%	\includegraphics[width=0.5\textwidth]{ospaone.eps}\\
%	\caption{{OSPA distance of order $p=1$ and cut-off $c=100$ over one trail.}}
%\end{figure}

\begin{figure}
	\centering
	\includegraphics[width=0.5\textwidth]{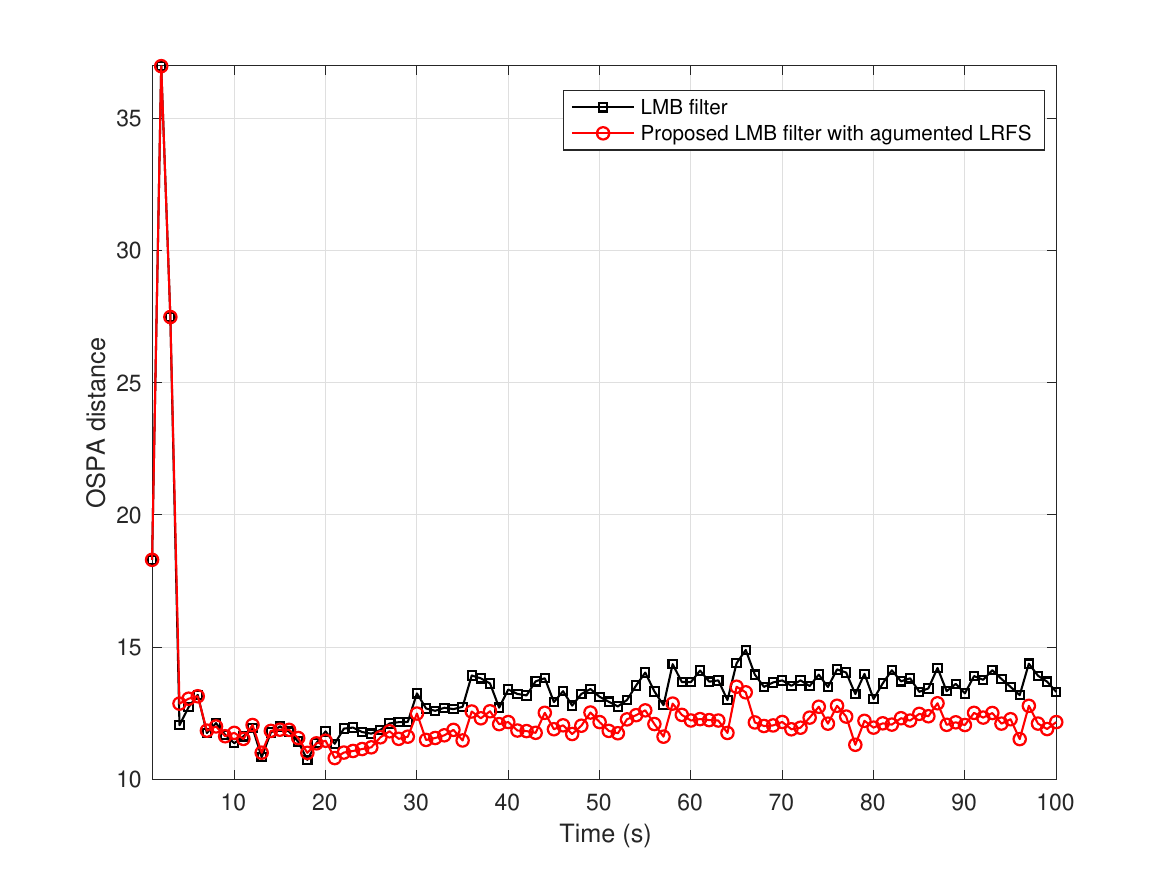}\\
	\caption{{Average OSPA distance  of order $p=1$ and cut-off $c=100$ over 100 MC trails.}}
\end{figure}

\begin{figure}
	\centering
	\includegraphics[width=0.5\textwidth]{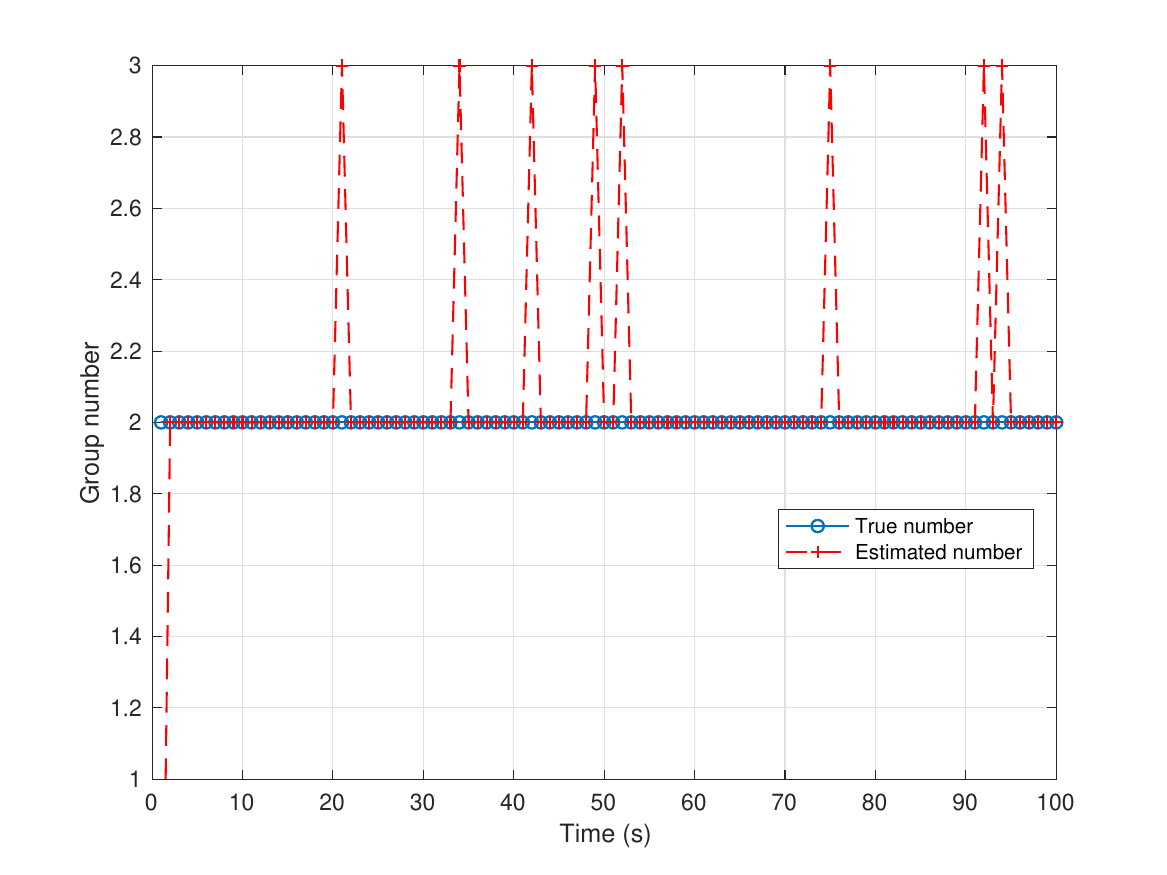}\\
	\caption{{Group number estimates.}}
\end{figure}

\section{Conclusion}\label{section:conclusion}
%%%%%%%%%%%%%%%%%%%%%%%%%%%%%%%%%%%%%%%%%%%%%%%%%%%%%%%%%%%%%%%%%%%%%%%%%%%%%%%%

This paper has addressed the GOT issue.
By integrating the group information of individual object into an  LRFS, 
a new kind of LRFS, named augmented \textcolor{blue}{LRFS, is developed.}
Furthermore, by the means of \textcolor{blue}{the augmented LRFS-based filter,}
the group information, along with the multi-object state and track labels,
have been propagated and updated in the tracking process.
In this way, the estimation of group structure and multi-object state  are integrated,
and  a holistic GOT \textcolor{blue}{framework  are established.}
At last, numerical  simulations have been presented to \textcolor{blue}{verify} the application  of the proposed augmented LRFS in the solving of the GOT issue.

\textcolor{blue}{Future work will be directed towards two asptects.  The first is to further explore the augmented LRFS theory, leading to more rigorous derivation and more detailed analysis for this theory. 
The second is to apply the augmented LRFS-based filter   to solve more complex tracking problems, such as the partially resolvable GOT  problem.}

\section*{ Acknowledgements}  
This work was supported in part by National Natural Science Foundation of China (62303109), 
Natural Science Foundation of Jiangsu Province of China (BK20230827), Start-Up Research Fund of Southeast University (RF1028623002), and  Zhishan Young Scholar Research Fund of Southeast University (2242024RCB0011).

\section*{Appendix}  
Suppose that the transition density of the group \textcolor{blue}{geometry} center follows
\begin{equation}\label{appendix1}
	f(c_{+}|c,l)=N(c_{+};Fc,Q),
\end{equation}
then (\ref{appendix1}) is equivalent to 
\begin{equation}\label{appendix2}
	c_k=Fc_{k-1}+v_k,
\end{equation}
where $v_k$ follows the Gaussian distribution with zero mean and covariance $Q$.
Assume that the evolution of group 
follows the  leader follower model, which  means that the deterministic state of any object is a translational offset of the \textcolor{blue}{geometry}
center (leader) of the group~\citep{zhang2022Seamless,Li20new},
then, we define  the offset as
\begin{equation}\label{appendix3}
	\bigtriangleup=x_k-c_{k}=x_{k-1}-c_{k-1},
\end{equation}
and we further have,
\begin{eqnarray}
	x_k&=&c_k+\bigtriangleup \\
	&=& Fc_{k-1}+v_k+x_{k-1}-c_{k-1} \\
	&=&(F-I)c_{k-1}++x_{k-1}+v_k. \label{appendix4}
\end{eqnarray}
Note that (\ref{appendix4}) is equivalent to  $f(x_{+}|x,c,l)=N(x_{+};(F-I)c+x,Q)$.
Thus, if the transition density of the group \textcolor{blue}{geometry} center follows $f(c_{+}|c)=N(c_{+};Fc,Q)$,
$f(x_{+}|x,c,l)=N(x_{+};(F-I)c+x,Q)$ holds true.

\bibliographystyle{elsarticle-num}        % Include this if you use bibtex 
\bibliography{new240704}    

%\begin{thebibliography}{00}
%%% Text of bibliographic item
%
%\bibitem{}
%
%\end{thebibliography}

\end{document}